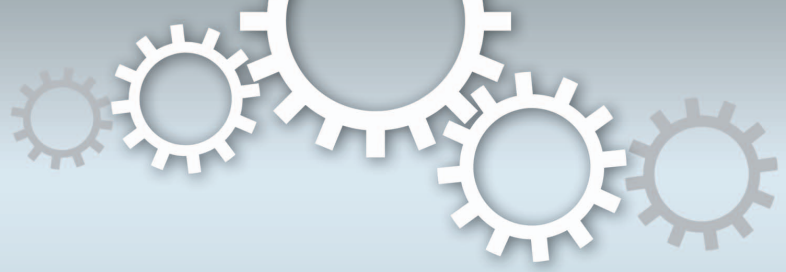

# SCIENTIFIC REPORTS

# Graphene, a material for high temperature devices – intrinsic carrier density, carrier drift velocity, and lattice energy

Yan Yin[1], Zengguang Cheng[2], Li Wang[1], Kuijuan Jin[1] & Wenzhong Wang[3]

[1]The Institute of Physics, Chinese Academy of Sciences, Beijing 100190, P. R. China, [2]National Center for Nanoscience and Technology, Beijing 100190, P. R. China, [3]Minzu University of China, Beijing 100081, P. R. China.

Heat has always been a killing matter for traditional semiconductor machines. The underlining physical reason is that the intrinsic carrier density of a device made from a traditional semiconductor material increases very fast with a rising temperature. Once reaching a temperature, the density surpasses the chemical doping or gating effect, any p-n junction or transistor made from the semiconductor will fail to function. Here, we measure the intrinsic Fermi level ($|E_F| = 2.93 \, k_B T$) or intrinsic carrier density ($n_{in} = 3.87 \times 10^6 \, cm^{-2} K^{-2} \cdot T^2$), carrier drift velocity, and G mode phonon energy of graphene devices and their temperature dependencies up to 2400 K. Our results show intrinsic carrier density of graphene is an order of magnitude less sensitive to temperature than those of Si or Ge, and reveal the great potentials of graphene as a material for high temperature devices. We also observe a linear decline of saturation drift velocity with increasing temperature, and identify the temperature coefficients of the intrinsic G mode phonon energy. Above knowledge is vital in understanding the physical phenomena of graphene under high power or high temperature.

S ince the discoveries of carbon nanotube and graphene[1,2], a significant amount of attention has been directed toward the room temperature and low temperature measurements in pursuing a high carrier mobility[3–5]. As people seek device applications for graphene, high field and high current measurements start to appear[6–10]. The importance of electron-phonon coupling is well recognized[11–13]. However, little attention has been made to the role of intrinsic carriers in graphene, because the great ability of graphene to tolerate the impact of temperature masks the existence of its intrinsic carriers. Thus, the importance of graphene as a material for high temperature devices is not noticed. Here we use a Raman method to measure the level of intrinsic carriers in a voltage biased graphene device up to 2400 K. We find the level of intrinsic carriers in graphene is an order of magnitude less sensitive to the temperature than that of traditional semiconductors, silicon and Germanium. The result tells us that a properly fabricated graphene p-n junction or transistor can operate well above 1500 K and will not fail until other parts melt. Working together with silicon carbide (SiC), graphene might be one of the building materials for coming high temperature devices. We also observe a linear decline of saturation drift velocity in graphene with increasing temperature in contrast to a constant saturation drift velocity in traditional semiconductors, as the result of the degeneracy of carriers and the Pauli exclusion principle. At the same time, we identify the temperature coefficients of the intrinsic G mode lattice vibration energy up to 2400 K. Above information highlights the great application potentials for graphene in high temperature devices, suggests a new exploring direction for graphene research, and is critical in the physics of graphene under high field, high current or high temperature conditions.

The graphene devices were fabricated by mechanical exfoliation from graphite. Figure 1a shows a sketch of the layout of graphene devices and electrical measurement. Contacts were formed on top of the graphene after steps of E-beam lithography, metal evaporation and liftoff. The contacts are 100 nm Au with 5 nm Cr between Au and graphene. The substrate is doped Si with 300 nm $SiO_2$ isolation layer on top. When an electrical bias is applied to a device, the drain electrode and Si substrate are grounded. A Kethley 2400 is used both as a voltage source and a current meter. During the measurements, the devices were kept in a flowing nitrogen gas environment in order to prevent the oxidization of the sample. A Horiba Raman microscope was used to measure the Raman spectra of the sample from the top. The excitation laser is a He-Ne 633 nm (1.96 eV) laser. The focusing objective is a long working distance 50× objective lens. Two graphene devices are measured; the optical pictures of these graphene devices are showing in Fig. 1b and 1c. Device A is 5 μm in length and 2.7 μm in width, and device B is 4.85 μm in length and 17.1 µm in width. For example, we measured the device A from $V_{sd} = 0$ V to $V_{sd} = 21$ V with an increasing step of 1 V, then a downward scan (with a step size of $-1$ V) back to $V_{sd} = 0$. For each different $V_{sd}$, a Raman spectrum from the center of the device was recorded; the currents of the device before and after each Raman measurement were recorded as well.

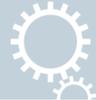

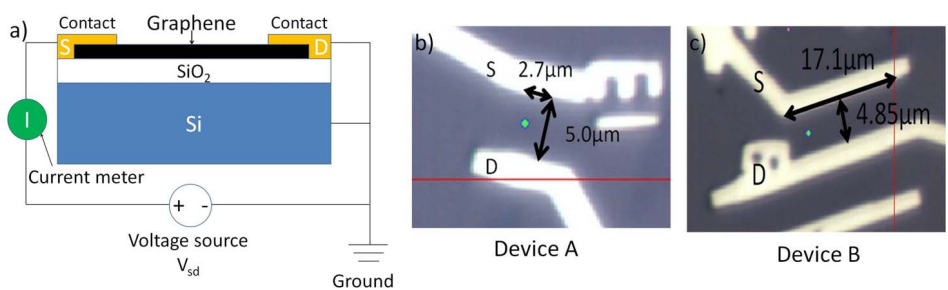

**Figure 1 | Graphene device information.** (a), The sketch of the stacking structure of the devices and the electrical measurement layout. (b), Optical picture of the graphene device A. (c), Optical picture of the graphene device B. The widths and lengths of the devices are marked in the pictures. The blue dots are the optical focus positions during the measurements of the devices. The letters D and S indicate the drain electrodes and source electrodes.

Our measurement result reveals the dominating impact of intrinsic carrier due to increasing temperature in the devices under high bias conditions. In Figure 2, we plot the measurement results of Raman spectra as a function of $V_{sd}$ for device A. We can see two main Raman peaks, G mode around 1600 cm$^{-1}$ and 2D mode[14] around 2600 cm$^{-1}$. In both subfigures, G mode intensity increases with a larger $V_{sd}$. At the same time, a dramatic increasing of the background signal appears with a high $V_{sd}$ voltage and this background will drop back down with the decreasing $V_{sd}$ during the downward scan. In an earlier report[15], C.-F. Chen et al. presented that the G mode intensity has resonant behavior when the twice of the Femi level in graphene matches the excitation photon energy or the photon energy minus a G mode phonon energy. They also reported an increasing background in the spectra near the resonant condition, and they attributed this high background signal to hot luminescence in graphene. Our data's resonant behaviors in the G mode intensity and background are very similar to those reported behaviors, therefore we believe that the G mode intensity change is due to a changing Femi level in the graphene device. The Femi level ($E_F$) in graphene is directly linked to the carrier density ($n$) with $n = \left(\frac{E_F}{\hbar v_F}\right)^2 / \pi$, where

$v_F$ is the graphene Fermi velocity, $\hbar$ is the reduced Planck constant. A higher Fermi level will have a higher carrier density, and vice versa. There are four possible reasons may change the device carrier density. They are intrinsic carrier density increasing due to rising temperature, permanent defects in graphene, back gating effect due to increasing $V_{sd}$, and the desorbing of the surface molecules. In comparison with Fig. 3a in C.-F. Chen et al.'s paper[15] after shifting the resonant peak due to different excitation photon energies (785 nm to 633 nm, 1.58 eV to 1.96 eV), we use the relative intensity of the G mode at $V_{sd} = 21$ V over base intensity at $V_{sd} = 0–5$ V, and determine that $|E_F| \approx 0.8$ eV or $n \approx 4 \times 10^{13}$ cm$^{-2}$ when $V_{sd} = 21$ V. First of all, this high carrier density clearly is not the result of permanent defects in the device, because the $V_{sd}$ downward scan shows this carrier density will go down with a lower voltage. For unintentional residual and surface adsorbate doping, the carrier density is normally at the scale of $10^{11}–10^{12}$ cm$^{-2}$, and this doping should only decrease with an increasing temperature. The back gating effect of the $V_{sd}$ can be estimated after calculating the back gating capacitance. For a 300 nm SiO$_2$ layer (permittivity $\varepsilon = 4$), the typical back gating capacitance $C_{bg} = 1.2 \times 10^{-8}$ F·cm$^{-2}$; for $V_{sd} = 20$ V, back gating carrier density at the middle of the device $n_{bg} \approx C_{bg}V_{sd}/2 = 7.5 \times 10^{11}$ cm$^{-2}$. This value is two orders of magnitude smaller than what we have ($n \approx 4 \times 10^{13}$ cm$^{-2}$). The only reasonable explanation for this high rising carrier density is the increasing intrinsic carrier density due to rising temperature within the device. Also, the scales of other effects, back gating and desorption of surface adsorbate, are a couple orders of magnitude smaller than the intrinsic carrier density under the high bias condition in our sample. We determine that the thermally generated intrinsic carrier is the dominating factor for those short samples under high bias voltage conditions.

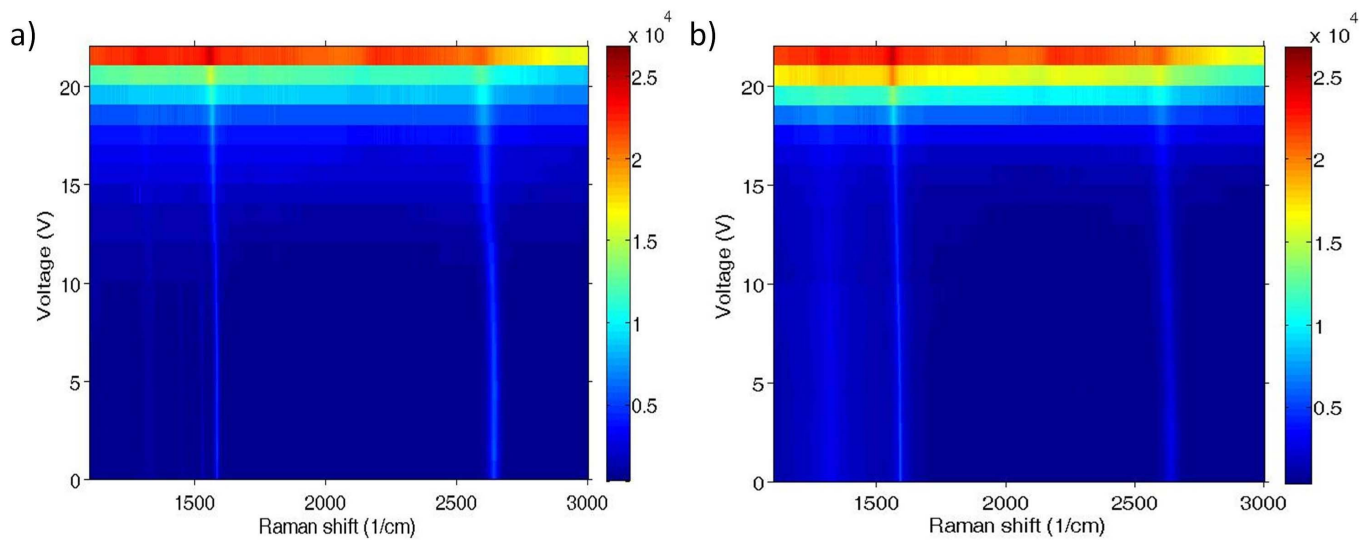

**Figure 2 | Raman spectra of device A plotted in color maps for different source drain voltages ($V_{sd}$).** (a), The spectra during the $V_{sd}$ upward scan. (b), The spectra during the $V_{sd}$ downward scan.

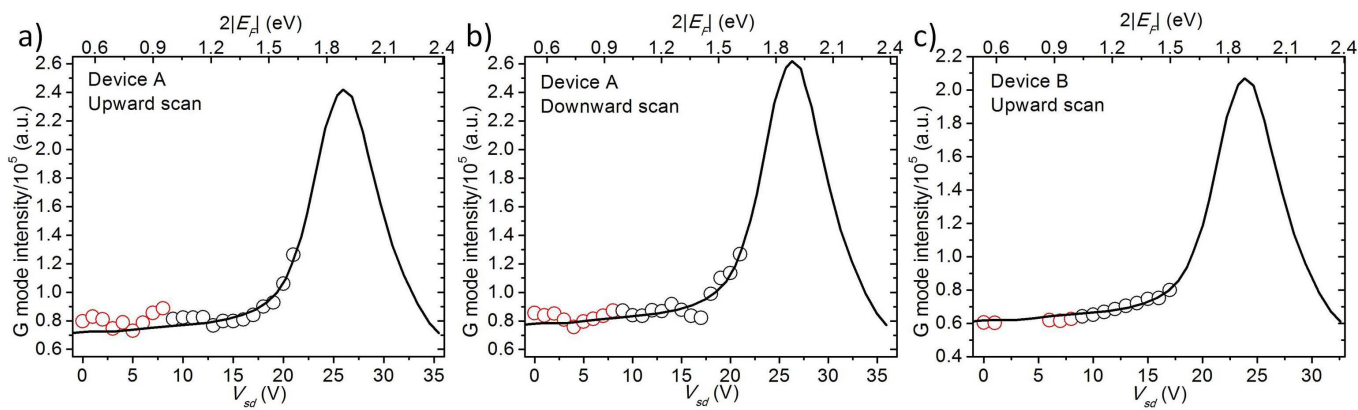

**Figure 3 | Measurement and fitting results for G mode intensities as a function of the $V_{sd}$.** (a), Device A during $V_{sd}$ upward scan. (b), Device A during $V_{sd}$ downward scan. (c), Device B during $V_{sd}$ upward scan. The data points are fitted with the curve from Fig. 3a in C.-F. Chen et al.'s paper[15] after shifting 0.38 eV (the difference between excitation photon energies, 785 nm to 633 nm or 1.58 eV to 1.96 eV). We only use the $V_{sd} \geq 9$ V data points for the fitting in order to give a better fitting weight for the high bias data points.

First, we will use the G mode intensity to determine the Fermi level in the devices. The Fermi level in the graphene device is changed with $V_{sd}$, but not directly through $V_{sd}$. The $V_{sd}$ drives the graphene device to generate heat; the heat generated carrier density rises in the device with an increasing temperature; with the help of external field from $V_{sd}$, heat generated carriers break balance between electron carriers and hole carriers, and the major carrier winning out raises the Fermi level in the graphene device. At the Dirac point and no external field, Tian Fang et al. calculated thermally generated carrier density[16], $n_{th} = \frac{\pi}{6} \left( \frac{k_B T}{h v_F} \right)^2$, where $k_B$ is the Boltzmann constant, T is the temperature. Their result gives out a $n_{th}$ for both the electrons and the holes, the electrons and holes cancel each other so the Fermi level is still zero. Clearly, our graphene devices under high field don't fit into their calculation's pre-condition, and their calculation result probably is numerically invalid for our case as well. In a connected real device with a certain electric voltage level, the Fermi level can go well above zero. For our devices, we assume the Fermi level due to rising intrinsic carrier density at the measurement point of a device,

$$|E_F| = \frac{\alpha}{2} V_{sd} + E_0,$$ (1)

where $\alpha$ is a device and location dependent constant, $E_0$ is a constant Fermi level when $V_{sd} = 0$. The reasons behind this assumption are: it is reasonable to guess $|E_F| \propto k_B T$, and we notice that device temperature $T$ has a linear relationship with $V_{sd}$. (This will be discussed later when we discuss transport data, and in supplementary information.) Equation (1) works much better for $V_{sd}$ downward scan than $V_{sd}$ upward scan. During $V_{sd}$ upward scan, the graphene device goes through desorption of surface adsorbate and increasing of permanent defects due to rising temperature[17], so the $E_0$ is not a constant. During $V_{sd}$ downward scan, the surface adsorbate was fully released at the upward scan step, and the permanent defects should be the same because the device is cooling, so the $E_0$ is about a constant during downward scan. This argument is supported by Figure 4a, which shows that the downward scan data for G mode positions shows a much better smoothness and data continuity than the upward scan data. In addition, Figure 4a shows that the G mode peak position, $\Omega_G = 1590.44$ cm$^{-1}$, for $V_{sd} = 0$ after the downward scan. I. Calizo et al. reported that the graphene G mode energy linear coefficient with temperature is $-0.016$ cm$^{-1}$/K for a temperature range 83 K–373 K, and $\Omega_G = 1584$ cm$^{-1}$ for $T = 0$ K[18], so $\Omega_G =$

1579.2 cm$^{-1}$ for $T = 298.15$ K (room temperature) if the Fermi level in graphene is zero. The blue shifted 11.24 cm$^{-1}$, in G mode energy at $V_{sd} = 0$ after downward scan, is due to the Fermi level $E_0$ in graphene. Because[15] $\Delta\Omega_G = |E_F| \times 42$ cm$^{-1}$eV$^{-1}$, we will have $E_0 = 0.2676$ eV for device A during downward scan. Figure 3 shows our measured G mode intensities as a function of the $V_{sd}$. The fitting curves are the curve retrieved from C.-F. Chen et al.'s paper[15] Fig. 3a for $I_G$ vs. $2|E_F|$, after blue shifted 0.38 eV. (They use a 785 nm–1.58 eV excitation laser and we use a 633 nm–1.96 eV excitation laser.) Our experimental data are fitted to the curve with two fitting parameters, $\alpha$ in equation (1) with constant $E_0 = 0.2676$ eV and a linear scaling factor between our G mode intensity and $I_G$ in C.-F. Chen et al.'s paper. The fitting is most accurate for device A during downward scan, because the $E_0$ value is measured for it. We do the same fitting for device A upward scan anyway, just to show a comparison. We only use the $V_{sd} \geq 9$ V data points for the fitting in order to give a better fitting weight for the high bias data points, which matter the most in this fitting. The G mode intensities at the high bias voltages already rise significantly above the random error range, especially for the data from device A. Therefore, our data have enough signal-to-noise ratios to make this fitting method valid, although our data doesn't cover the full resonant peak. For device A, the upward scan data fitting gives $\alpha = 0.0520$ eV/V; the downward scan data fitting gives $\alpha = 0.0513$ eV/V; these two values are close and we take the downward scan **$\alpha = 0.0513$ eV/V** as the value for device A. Data from device B has lower quality than the data from device A, however we use device B for the verification purpose only. Because device B broke down after the measurement at $V_{sd} = 17$ V during the upward scan, no data is available for device B $V_{sd}$ downward scan and we can't accurately determine the $E_0$. In Fig. 4a, the high bias ($V_{sd} \geq 6$ V) data points' behaviors are similar between the upward scan and downward scan for device A, this might suggest that the average $E_0$ value for high bias conditions in upward scan is close to the $E_0$ value in downward scan. We also consider device B has a similar length and goes through similar processes as device A, it is likely that device B has a similar defect level as device A. So, we take $E_0 = 0.2676$ eV for device A, as an approximation value, for the fitting of device B high bias data in Fig. 3c. The fitting gives $\alpha \approx 0.0566$ eV/V for device B.

Next, we will find out the temperatures at the measurement point for difference $V_{sd}$ situations using G mode peak position information. The key information we used to determine temperature is: the value of room temperature, the intrinsic G mode peak position shift (after correcting the doping effect) as a result of temperature change, the reported G mode energy temperature coefficient for temperature range 83 K to 373 K from Ref. 18, and a 2nd order polynomial fitting of our G mode peak position. The 2nd order fitting gives the relation-ship between the first and second temperature coefficients; our device temperature range overlaps with Ref. 18, so the reported the linear temperature coefficient from Ref. 18 sets another condition for the first and second temperature coefficients. Combining both conditions, we can have the values of the first and second temperature coefficients for G mode energy. With the value of room temperature and G mode Raman shift of each data point, we can have the device temperature of each data point. Figure 4 shows the G mode peak positions as a function of the $V_{sd}$ for device A and device B. In comparison between upward scan and downward scan G mode positions for device A, the downward scan data shows a much better smoothness and data continuity. The surface adsorbate was fully released during upward scan, so no chemical doping changes during downward scan. The device goes through cooling during downward scan, the permanent defects have a constant amount, and only shift the downward scan data a constant amount in G mode energy. Therefore, we use the downward scan data for the further data analyzing. During $V_{sd}$ downward scan, the G mode peak position, $\Omega_G$, satisfies $\Omega_G = \Omega_{G\_T} + \Delta\Omega_{G\_in} + \Delta\Omega_{G\_defect}$, where $\Omega_{G\_T}$ is the intrinsic G mode energy as a function of the temperature, $\Delta\Omega_{G\_in}$ is the G mode energy shift due to rising Fermi level (intrinsic carrier density) as a function of $V_{sd}$, $\Delta\Omega_{G\_defect}$ is a constant from the permanent defects level in the graphene. We know that $\Delta\Omega_{G\_in} + \Delta\Omega_{G\_defect} = |E_F| \times 42$ cm$^{-1}$eV$^{-1}$ and $|E_F| = 0.0513 \frac{eV}{V} \cdot \frac{V_{sd}}{2} + 0.2676$ eV for device A, then we can have a corrected G mode energy, $\Omega_{G\_T} = \Omega_G - (\Delta\Omega_{G\_in} + \Delta\Omega_{G\_defect})$, and we plot the new values in Fig. 4a as blue circle points. The new, after correction, G mode energies are fitted very well with a $2^{nd}$ order polynomial function.

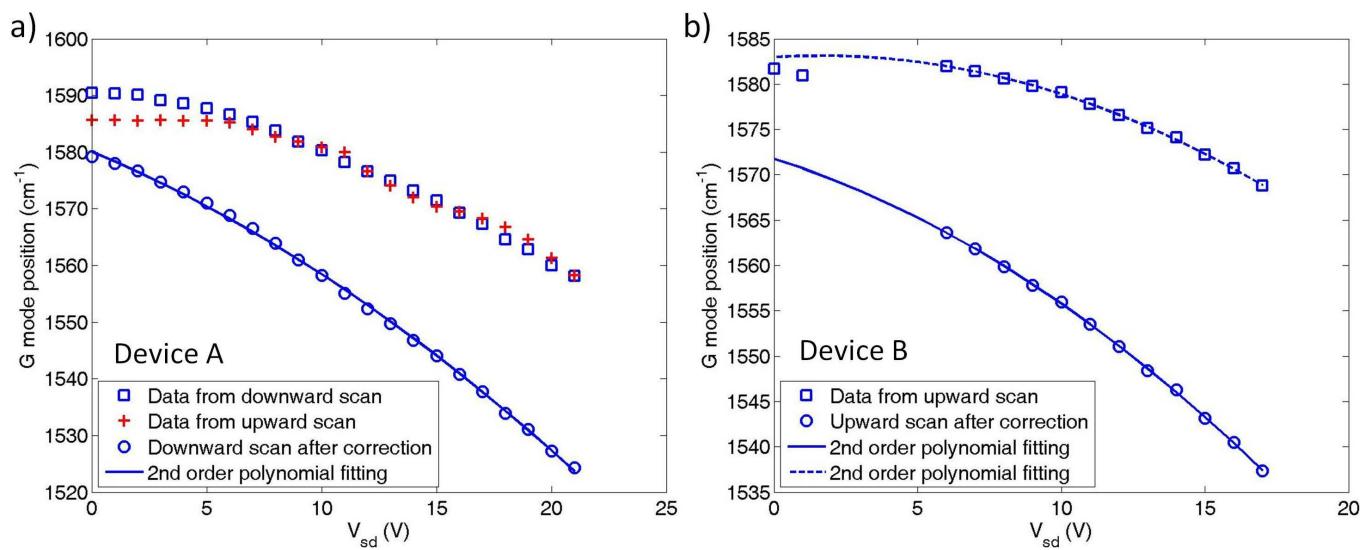

**Figure 4 | The G mode peak positions as a function of the $V_{sd}$.** (a), Device A. The blue square, red cross, and blue circle data points are the experimental data points for $V_{sd}$ downward scan, $V_{sd}$ upward scan, and the G mode position for downward scan after taking out the effect of carrier density, respectively. The blue curve is the fitting result for the corrected G mode position with a $2^{nd}$ order polynomial function. (b), Device B. The blue square and blue circle data points are the experimental data points for $V_{sd}$ upward scan and the G mode position after taking out the effect of carrier density. The blue solid curve and dash curve are the fitting results for the corrected G mode position and the original data points, respectively, with a $2^{nd}$ order polynomial function. Only the data points of $V_{sd} \geq 6$ V are used for the fittings of device B.

$$\Omega_{G\_T} = a_1 V_{sd}^2 + a_2 V_{sd} + a_3, \qquad (2)$$

where $a_1$, $a_2$ and $a_3$ are fitting parameters. For device A, we get $a_1 = -0.0474$ cm$^{-1}$V$^{-2}$, $a_2 = -1.6901$ cm$^{-1}$V$^{-1}$, and $a_3 = 1580$ cm$^{-1}$. We have not found any report about the coefficients of the graphene G mode energy as a function of the temperature for a temperature range up to 2500 K. H. Herchen and M. Cappelli reported that a first order Raman mode energy of diamond has a $2^{nd}$ order polynomial dependence on temperature up to 2000 K[19]. So, it is reasonable to assume, for G mode in graphene, that

$$\Omega_{G\_T} = b_1 T^2 + b_2 T + b_3, \qquad (3)$$

where $b_1$, $b_2$ and $b_3$ are graphene dependent only constants. The temperature of the device appears a linear relationship with $V_{sd}$, and we define

$$T = \beta V_{sd} + T_{RT}, \qquad (4)$$

where $\beta$ is a device and location dependent constant and $T_{RT}$ is the room temperature, 298.15 K. Combining equations (2), (3) and (4), we will have

$$a_1 = b_1 \beta^2, \ a_2 = 2\beta T_{RT} b_1 + b_2 \beta. \qquad (5)$$

I. Calizo et al. reported that the graphene G mode energy linear coefficient with temperature is $-0.016$ cm$^{-1}$/K for a temperature range 83 K–373 K, and $\Omega_{G\_T} = 1584$ cm$^{-1}$ for $T = 0$ K[18]. Because $b_1$, $b_2$ and $b_3$ are graphene dependent only constants, we have

$$\left.\frac{\Delta\Omega_{G\_T}}{\Delta T}\right|_{T = 228K} = 2b_1 T|_{T = 228K} + b_2 = -0.016 \ cm^{-1}/K,$$

$$b_3 = 1584 \ cm^{-1}, \qquad (6)$$

in which 228 K is the average of 83 K and 373 K. Combining equations (5) and (6) with knowing values for $a_1$ and $a_2$ from fitting of device A, we can solve $b_1 = -4.596 \times 10^{-6} \ cm^{-1}K^{-2}$, $b_2 = -0.0139 \ cm^{-1}K^{-1}$, $b_3 = 1584 \ cm^{-1}$, and $\beta = 101.5 \ K/V$ for the measurement point in device A. Finally, we combine equation (1) and (4), and have $|E_F| = \gamma T$ to describe a graphene device intrinsic Fermi level as a function of the temperature, where $\gamma = \alpha/2\beta$. From device A, our measurement gives out a graphene device intrinsic Fermi level, $|E_F| = 2.53 \times 10^{-4} \ eVK^{-1} \cdot T$ or $|E_F| = 2.93 \ k_BT$. Using graphene $\hbar v_F = 1.16 \times 10^{-28} \ J \cdot m$, we have a graphene device intrinsic carrier density, $n_{in} = 3.87 \times 10^6 \ cm^{-2}K^{-2} \cdot T^2$.

In Fig. 4b, the square points are the measured G mode positions of device B as a function of $V_{sd}$. The circle points are the corrected G mode energies using the same correction method for device A, but with $|E_F| = 0.0565 \frac{eV}{V} \cdot \frac{V_{sd}}{2} + 0.2676$ eV. For the $\beta$ value, it is invalid to process the limited data from device B with the same method for device A. Luckily, $a_1$ is a fitting parameter independent from the values of α and $E_0$. In order to prove this point, we did the 2nd order polynomial fittings for both measured G mode positions and corrected G mode energies, and plotted the fitting curves as a dash line and a solid line, respectively, in Fig 4b. Both fittings give $a_1 = -0.0603$ cm$^{-1}$V$^{-2}$. Because $b_1$ is a graphene dependent only constant and should be same for all graphene, we can use the $b_1 = -4.596 \times 10^{-6}$ cm$^{-1}$K$^{-2}$ value from device A, for device B. Using equation (5), $a_1 = b_1\beta^2$, we can get $\beta = 114.5$ K/V for device B. Together with α ≈ 0.0566 eV/V for device B, we get γ = 2.47 × $10^{-4}$ eV/K from device B. This γ value of device B is consistent with the γ value from device A. Such an agreement between device A and device B supports the validity of our method.

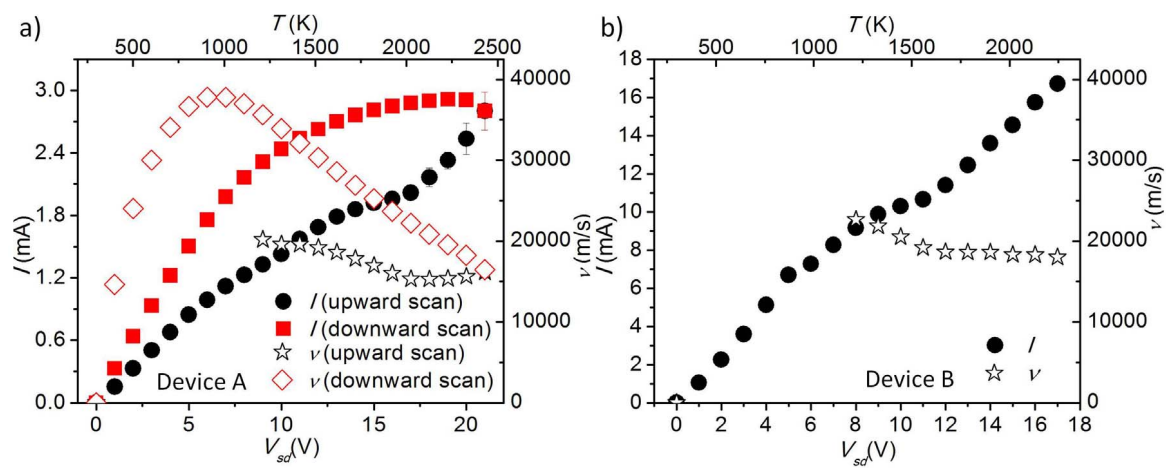

**Figure 5 | Device I–V curves, temperatures, and carrier drift velocities.** (a), Device A, the defect level is $4 \times 10^{12}$ cm$^{-2}$. (b), Device B. Red solid square points and black solid round points are the I–V data during $V_{sd}$ downward and upward scan, respectively, using left y axis. The current values are the averages of the currents before and after the Raman measurements, the differences of the two are plotted as error bars. Red hollow diamond points and black hollow star points are carrier drift velocities during $V_{sd}$ downward and upward scan, respectively, using right y axis. Only the data points with carrier density $n \geq 1.4 \times 10^{17}$ m$^{-2}$ during the upward scan are plotted in order to control the error in drift velocities due to the uncertainty in $E_0$. The top axis plots the device temperature, which is calculated from $V_{sd}$ using equation (4).

Graphene shows a great potential as a material for high temperature devices, compared to traditional semiconductors. High temperature devices are demanded for military, astronautic, and heavy industry applications, where high power density is required or high temperature environment can't be avoided. In order to make a nonlinear semiconductor device (p-n junction or transistor) functioning, the chemically doped carriers or the gating generated carriers have to be the controlling factor within the semiconductor material. However, when the device temperature increases, the semiconductor's intrinsic carrier density will increase. Once the temperature reaches a certain point that, the semiconductor's intrinsic carrier surpasses the chemical doping or gating effect and dominates the device's response, the nonlinear device will simply fail its function. For instance, a Si device's intrinsic carrier density will double if temperature increases 8 K from room temperature, the maximum operation temperature for a Si device is 250°C; a Ge device's intrinsic carrier density will double if temperature increases 12 K from room temperature, the maximum operation temperature for a Ge device is 100 °C[20]. Normally for a semiconductor, a larger band gap gives a higher operation temperature; however, a larger band gap also increases the difficulty of carrier density tuning. Graphene is semimetallic, not a semiconductor. By a doping or field effect, graphene can be made into a p-n junction or a transistor as well. Running our numbers for graphene, we find that it will take 125 K for graphene intrinsic carriers to double from room temperature. At 1500 K, graphene intrinsic carrier density is $8.7 \times 10^{12}$ cm$^{-2}$. Multiple research reports[15,21,22] have demonstrated that a top gating method or a chemical doping can tune the graphene carrier density to a level above ±3 × $10^{13}$ cm$^{-2}$. Therefore, we foresee that a well-fabricated top gated graphene transistor or a chemical doped graphene p-n junction will be still well functioning at a temperature above 1500 K. A lot of things melt at 1500 K, for example gold melting temperature is about 1340 K. So, such a graphene transistor or graphene p-n junction will never fail due to rising intrinsic carrier density before other parts start to melt. Above analyses reveal the great potential of graphene as the material for high temperature devices and point out a new research direction for graphene applications.

Finally, we will show and discuss carrier drift velocity in graphene up to 2400 K. Fig. 5 shows the device I–V curves, temperatures and carrier drift velocities of device A and device B. Carrier drift velocity, $v$, equates to $I/(qWn)$, where $I$ is the current, $q$ is the electron charge, $W$ is the device width, $n$ is the carrier density. The device B broke down after $V_{sd} > 17$ V and the device A was also on the brink of breakdown at $V_{sd} = 21$ V and showed clear runaway effect, in which the current keeps increasing and feeds itself with a constant driving $V_{sd}$. The maximum temperature is about 2400 K for device A and 2200 K for device B, those temperatures are consistent with the theoretically calculated graphene breakdown temperature, 2230 ± 700 K, in vacuum[8]. The drift velocities from device A, $V_{sd}$ downward have the best accuracy compared to the values from upward scans, and map out a very nice three stage behavior under high field. In the first stage, the drift velocity increases linearly with the field, Ohm's law holds perfectly. Then with a further increased field, lattice scattering dominates, the drift velocity still increases but with a reducing rate, Ohm's law still works but with a reducing conductivity. In the last stage, after the field reaches a certain level, intrinsic carrier dominates, the drift velocity stops to increase. Traditionally, the drift velocity at this point is called the saturation drift velocity, and this saturation drift velocity stays the same for traditional semiconductors if the field increases further. For example, for Si, it is about 1 × $10^5$ m/s. Our graphene sample shows a linear drop of "saturation" drift velocity with the increasing temperature. This reducing saturation drift velocity of graphene has been predicted by theoretical calculation considering the degeneracy of carriers and the Pauli exclusion principle[23]. The same calculation gives a saturation drift velocity value, 2 × $10^5$ m/s, for an ideal graphene with a doping level $10^{13}$ cm$^{-2}$ at 300 K. The maximum drift velocity of our graphene with a defect level of $4 \times 10^{12}$ cm$^{-2}$ is about 4 × $10^4$ m/s at 1000 K. Our value is not far from the calculation value after considering the temperature difference and the scattering from the defects. The declining saturation drift velocity in graphene helps delay the current runaway point, where the current keeps increasing on its own with a constant driving voltage, to an even higher temperature; but it also limits the ability of graphene in delivering large current. Another thing we want to point out is that Ohm's law fails totally in the intrinsic carrier dominated third stage. The electrons and the lattice are in a non-equilibrium state, in which the energy electrons taking from the field is not fully dumped to the lattice. This is why the device temperature has a linear relationship with V, not the IV. The difference in drift velocity for a same $V_{sd}$ between downward and upward scan is due to different effective electron temperatures, despite a similar lattice temperature. During downward scan a graphene device appears a smaller contact resistant (or barrier) than the upward scan[24], therefore, downward scan data has a larger current, electrons carrying more energy and moving faster, or a higher effective electron temperature than the upward scan data.

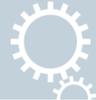

In summary, we use Raman method to quantitatively measure the graphene intrinsic carrier density, drift velocity and G mode phonon energy as functions of temperature up to 2500 K. We find that the intrinsic carrier density of graphene is an order of magnitude less sensitive to temperature than that of traditional semiconductor materials. We foresee graphene becomes the material for high temperature devices and a properly fabricated graphene transistor or p-n junction will not fail at all before other parts melt. We map out the three stage behavior of carrier drift velocity as a function of the field or temperature, and graphene shows a linear decline in "saturation" drift velocity in contrast to a constant saturation drift velocity in traditional semiconductor materials. We also find the G mode phonon intrinsic energy coefficients with temperature, $\Omega_G = -4.596 \times 10^{-6} \ cm^{-1}K^{-2}T^2 + -0.0139 \ cm^{-1}K^{-1}T + 1584 \ cm^{-1}$. All of above reveal the great potential of graphene as a foundational material for high temperature devices, identify a new research direction for graphene applications, and are vital in understanding the physics of graphene under high field or high temperature.

## Acknowledgments

This work was supported by the National Basic Research Program of China (No. 2014CB921001 and No. 2014CB339800), by the Young Scientists Fund of the National Natural Science Foundation of China (Grant No. 11004231), by the National Natural Science Foundation of China (Grant No. 21161120321), and by the Scientific Research Start-up Fund for the Returned Overseas Chinese Scholars from Ministry of Education of China. Y.Y. also acknowledges Ying Fang's good suggestions for the experiments, Yuping Yang's help in arranging the Raman instrument, and helpful discussions with Wenxin Wang.

## Author Contributions

Y.Y. designed the experiments, carried out the measurements, performed the data analyses and drafted the initial manuscript; Z.C. contributed to sample fabrication; L.W. and K.J. provided the instruments for transport measurements and funding support, W.W. provided the Raman instrument. All authors discussed the results and wrote the paper together.

## Additional information

**Supplementary information** accompanies this paper at http://www.nature.com/scientificreports

**Competing financial interests:** The authors declare no competing financial interests.

**How to cite this article:** Yin, Y., Cheng, Z., Wang, L., Jin, K. & Wang, W. Graphene, a material for high temperature devices – intrinsic carrier density, carrier drift velocity, and lattice energy. *Sci. Rep.* **4**, 5758; DOI:10.1038/srep05758 (2014).